\documentclass[12pt]{article}

\usepackage{times}
\usepackage{fullpage}

\usepackage{amsfonts}
\usepackage{amssymb}
\usepackage{amsmath}
\usepackage{latexsym}
\usepackage{url}
\usepackage[breaklinks]{hyperref}

\newcommand{\R}{\mathbb{R}}
\newcommand{\ADV}{\mathrm{ADV}}
\newcommand{\adv}{\mathrm{ADV}}
\newcommand{\AND}{\mathrm{AND}}
\newcommand{\SA}{\mathrm{SA}}
\newcommand{\MM}{\mathrm{MM}}

\newcommand{\Gf}{{\Gamma_{\!f}}}

\newcommand{\df}{{\delta_f}}

\newcommand{\Gg}[1]{\Gamma_{\!g_{#1}}}

\newcommand{\dg}[1]{{\delta_{g_{#1}}}}
\newcommand{\dgb}[2]{{\delta_{g_{#1}}^{\upharpoonright #2}}}

\newcommand{\Gh}{{\Gamma_{\!h}}}
\renewcommand{\dh}{{\delta_h}}
\newcommand{\tx}{{\tilde x}}
\newcommand{\ty}{{\tilde y}}

\newcommand{\sublam}[2]{{\|#1 \circ D_{#2}\|}}
\newcommand{\fxnefy}[1]{{x,y \atop #1(x) \ne #1(y)}}
\newcommand{\xineyi}[1]{{#1: x_{#1} \ne y_{#1}}}
\newcommand{\innersum}{\genfrac{}{}{0pt}{}{x,y}{x\in X_a,y\in X_b}}
\newcommand{\advmax}{\genfrac{}{}{0pt}{}{\Gamma \ge 0}{\Gamma \ne 0}}

\long\def\rem#1{}
\def\01{\{0,1\}}

\newcommand{\norm}[1]{\| #1 \|}
\newcommand{\sn}[1]{\| #1 \|}

\newtheorem{definition}{Definition}

\newtheorem{theorem}{Theorem}
\newtheorem{lemma}[theorem]{Lemma}
\newtheorem{corollary}[theorem]{Corollary}

\newcommand{\thmref}[1]{\hyperref[#1]{{Theorem~\ref*{#1}}}}
\newcommand{\lemref}[1]{\hyperref[#1]{{Lemma~\ref*{#1}}}}
\newcommand{\corref}[1]{\hyperref[#1]{{Corollary~\ref*{#1}}}}
\newcommand{\eqnref}[1]{\hyperref[#1]{{Equation~(\ref*{#1})}}}
\newcommand{\claimref}[1]{\hyperref[#1]{{Claim~\ref*{#1}}}}
\newcommand{\remarkref}[1]{\hyperref[#1]{{Remark~\ref*{#1}}}}
\newcommand{\propref}[1]{\hyperref[#1]{{Proposition~\ref*{#1}}}}
\newcommand{\factref}[1]{\hyperref[#1]{{Fact~\ref*{#1}}}}
\newcommand{\defref}[1]{\hyperref[#1]{{Definition~\ref*{#1}}}}
\newcommand{\exampleref}[1]{\hyperref[#1]{{Example~\ref*{#1}}}}
\newcommand{\hypref}[1]{\hyperref[#1]{{Hypothesis~\ref*{#1}}}}
\newcommand{\secref}[1]{\hyperref[#1]{{Section~\ref*{#1}}}}
\newcommand{\chapref}[1]{\hyperref[#1]{{Chapter~\ref*{#1}}}}
\newcommand{\apref}[1]{\hyperref[#1]{{Appendix~\ref*{#1}}}}

\newcommand{\ignore}[1]{}

\newenvironment{proof}[1][Proof.]{
	\par
	\noindent \textbf{#1}
}{
	\unskip
	\nobreak\hfill\penalty50\hskip3pt\hbox{}\nobreak\hfill
	\hbox{$\Box$}\par\bigskip
}

\begin{document}
\title{Tight adversary bounds for composite functions}
\author{%
  Peter H\o yer%
  \thanks{Department of Computer Science,
  University of Calgary.
  Supported by Canada's Natural Sciences and Engineering 
  Research Council (NSERC),
  the Canadian Institute for Advanced Research (CIAR), 
  and The Mathematics of Information Technology and Complex 
  Systems (MITACS).}\\
  {\tt hoyer@cpsc.ucalgary.ca}
\and
  Troy Lee%
  \thanks{%
  LRI, Universit\'e Paris-Sud.  Supported by a Rubicon grant from 
  the Netherlands Organisation for Scientific Research (NWO).  Part of this 
  work conducted while at CWI, Amsterdam.}\\
  {\tt lee@lri.fr}
\and
  Robert \v Spalek%
  \thanks{%
  CWI and University of Amsterdam.  
  Supported in part by the EU fifth framework project RESQ, 
  IST-2001-37559.  
  Work conducted in part while visiting the University of Calgary.}\\
  {\tt sr@cwi.nl}
}
\date{}
\maketitle

\begin{abstract}
The quantum adversary method is a versatile method for proving
lower bounds on quantum algorithms.  It yields tight bounds for many
computational problems, is robust in having many equivalent
formulations, and has natural connections to classical lower bounds.  
A further nice property of the adversary method is that it behaves very
well with respect to composition of functions.  We generalize the adversary
method to include costs---each bit of the input can be given an 
arbitrary positive cost representing the difficulty of querying that bit.
We use this generalization to exactly capture the adversary
bound of a composite function in terms of the adversary bounds of its 
component functions.  Our results generalize and unify previously known 
composition properties of adversary methods, and yield as a simple corollary 
the $\Omega(\sqrt{n})$ bound of Barnum and Saks on the quantum query complexity
of read-once functions.
\end{abstract}

\section{Introduction}
One of the most successful methods for proving lower bounds on quantum
query complexity is via adversary arguments.  The basic idea behind 
the adversary method is that if a query algorithm successfully computes 
a Boolean function $f$, then in particular it is able to 
``distinguish'' $0$-inputs from $1$-inputs.  There are many different ways
to formulate the progress an algorithm makes in distinguishing $0$-inputs 
from $1$-inputs by making queries --- these varying formulations have led to 
several versions of the adversary method including Ambainis' original 
weight schemes \cite{Amb02,Amb03}, the 
Kolmogorov complexity method of Laplante and Magniez \cite{LM04}, and a 
bound in terms of the matrix spectral norm due to 
Barnum, Saks and Szegedy \cite{BSS03}.  Using the duality theory of 
semidefinite programming, \v{S}palek and Szegedy \cite{SS06} show that in fact
all of these formulations are equivalent.

We will primarily use the spectral formulation of the adversary method.  
Let $Q_2(f)$ denote the two-sided bounded-error query 
complexity of a Boolean function $f: S \rightarrow \01$, with 
$S \subseteq \01^n$.  Let $\Gamma$ be a symmetric matrix with rows and columns 
labeled by elements of $S$.
We say that $\Gamma$ is an \emph{adversary matrix for $f$} if 
$\Gamma[x,y]=0$ whenever $f(x)=f(y)$.  The spectral adversary method states 
that $Q_2(f)$ is lower bounded by a quantity $\ADV(f)$ defined in terms
of~$\Gamma$.
\begin{theorem}[\cite{BSS03}]
For any function $f: S \rightarrow \01$, with $S \subseteq \01^n$ and any 
adversary matrix $\Gamma$ for~$f$, let
\begin{equation*}
\ADV(f) = \max_{\advmax}\  
  \frac{\sn{\Gamma}}{\max_i \sn{\Gamma \circ D_i}}.
\end{equation*}
Then $Q_2(f)=\Omega(\adv(f))$.
\end{theorem}
Here $D_i$ is the zero-one valued matrix defined by $D_i[x,y]=1$ if
and only if bitstrings $x$ and $y$ differ in the $i$-th coordinate,
and $\sn{M}$ denotes the spectral norm of the matrix~$M$.

One nice property of the adversary method is that it behaves very 
well for iterated functions.  For a function $f:\01^n \rightarrow \01$ we 
define the $d$-th iteration of $f$ recursively as
$f^1 = f$ and $f^{d+1} = f \circ (f^d, \dots, f^d)$ for $d \ge 1$.
Ambainis~\cite{Amb03} shows that if $\ADV(f) \geq a$ then $\ADV(f^d) \geq a^d$.
Thus by proving a good adversary bound on the base function $f$, one can 
easily obtain good lower bounds on the iterates of $f$.
In this way, Ambainis shows a super-linear gap between the 
bound given by the polynomial degree of a function and the adversary 
method, thus separating polynomial degree and quantum query complexity.

Laplante, Lee, and Szegedy~\cite{LLS06} show 
a matching upper bound for iterated functions, namely that if $\ADV(f) \leq a$
then $\ADV(f^d) \leq a^d$.  
Thus we conclude that the adversary method possesses the following composition
property.
\begin{theorem}[\cite{Amb03,LLS06}]
\label{thm:iterated}
For any function $f:S \rightarrow \01$, with $S \subseteq \01^n$ and natural 
number $d>0$,
\begin{equation*}
\ADV(f^d) = \ADV(f)^d.
\end{equation*}
\end{theorem}

A~natural possible generalization of Theorem~\ref{thm:iterated} is to
consider composed functions that can be written in the form
\begin{equation}\label{eq:composed}
h = f \circ (g_1, \dots, g_k).
\end{equation}
One may think of $h$ as a two-level decision tree with the top node
being labeled by a function $f:\01^k \rightarrow \01$, and each of
the $k$ internal nodes at the bottom level being labelled by a
function $g_i:\01^{n_i} \rightarrow \01$.  We do not require that the
inputs to the inner functions $g_i$ have the same length.  An~input $x
\in \01^n$ to $h$ is a bit string of length $n = \sum_i n_i$, which we
think of as being comprised of $k$ parts, $x=(x^1, x^2, \ldots, x^k)$,
where $x^i \in \01^{n_i}$.  We may evaluate $h$ on input $x$ by first
computing the $k$ bits $\tx_i = g_i(x^i)$, and then evaluating $f$ on
input $\tx = (\tx_1, \tx_2, \ldots, \tx_k)$.

It is plausible, and not too difficult to prove, that if $a_1 \leq
\ADV(f) \leq a_2$ and $b_1 \leq \ADV(g_i) \leq b_2$ for all~$i$, then
$a_1 b_1 \leq \ADV(h) \leq a_2 b_2$.  In particular, if the adversary
bounds of all of the sub-functions $g_i$ are equal 
(i.e., $\ADV(g_i)=\ADV(g_j)$ for all $i,j$), then we can give an exact 
expression for the adversary bound on $h$ in terms of the adversary bounds of 
its sub-functions,
\begin{equation}\label{eq:allequal}
\ADV(h) = \ADV(f) \cdot\ADV(g_i),
\end{equation}
It is not so clear, however, what the exact adversary bound of $h$ should be 
when the adversary bounds of the sub-functions $g_i$ differ.  The purpose of 
this paper is to give such an expression.

To do so, we develop as an intermediate step a new generalization of
the adversary method by allowing input bits to be given an 
arbitrary positive cost.  For any function $f: \01^n \rightarrow \01$,
and any vector $\alpha \in \R_{+}^n$ of length $n$ of positive
reals, we define a quantity $\ADV_\alpha(f)$ as follows:  
\[
\ADV_\alpha(f) = \max_{\advmax} \min_i \left\{ \alpha_i
  {\sn{\Gamma} \over \sn{\Gamma \circ D_i}} \right\}.
\]

One may think of $\alpha_i$ as expressing the cost of querying the $i$-th 
input bit $x_i$.  For example, $x_i$ could be equal to the parity of 
$2 \alpha_i$ new input bits, or, 
alternatively, each query to $x_i$ could reveal only a fraction of 
$1/\alpha_i$ bits of information about~$x_i$.  When $\alpha=(a,\ldots,a)$ 
and all costs are equal to $a$, the new adversary bound $\ADV_\alpha(f)$ 
reduces to $a \cdot \ADV(f)$, the product of $a$ and the standard adversary
bound $\ADV(f)$.  In particular, when all costs $a=1$, we have the 
original adversary bound, and so $Q_2(f)=\Omega(\ADV_{\vec 1}(f))$.
When $\alpha$ is not the all one vector, then $\ADV_{\alpha}(f)$ will not 
necessarily be a lower bound on the quantum query complexity of $f$, but 
this quantity can still be very useful in computing the adversary bound of
composed functions, as can be seen in our main theorem:

\begin{theorem}
[Exact expression for the adversary bound of composed functions] 
\label{thm:composition} For any
function $h: S \rightarrow \01$ of the form $h = 
f \circ (g_1, \dots, g_k)$ with domain $S \subseteq \01^n$, 
and any cost function $\alpha \in \R_{+}^n$,
\begin{equation*}
\ADV_\alpha(h) = \ADV_\beta(f),
\end{equation*}
where $\beta_i = \ADV_{\alpha^i}(g_i)$,
$\alpha = (\alpha^1, \alpha^2, \ldots, \alpha^k)$,
and $\beta = (\beta_1, \ldots, \beta_k)$.
\end{theorem}

The usefulness of this theorem is that it allows one to divide and conquer ---
it reduces the computation of the adversary bound for $h$ into the disjoint
subproblems of first computing the adversary bound for each $g_i$, and then, 
having determined $\beta_i=\ADV(g_i)$, computing
$\ADV_\beta(f)$, the adversary bound for $f$ with costs $\beta$.

One need not compute exactly the adversary bound for each $g_i$ to apply the
theorem.  Indeed, a bound of the form $a \le \ADV(g_i) \le b$ for 
all $i$ already gives some information about $h$.
\begin{corollary}
If $h=f \circ (g_1, \ldots, g_k)$ and $a \le \ADV(g_i) \le b$ for all $i$,
    then $a \cdot \ADV(f) \le \ADV(h) \le b \cdot \ADV(f)$.
\end{corollary}

One limitation of our theorem is that we require the 
sub-functions $g_i$ to act on disjoint subsets of the input bits.  Thus one
cannot use this theorem to compute the adversary bound of any function by, 
say, proceeding inductively on the structure of a 
$\{\wedge,\vee,\neg\}$-formula 
for the function.  One general situation where the theorem can be applied,
however, is to read-once functions, as by definition these functions are 
described by a formula over $\{\wedge, \vee, \neg\}$ where each variable 
appears only once.  To demonstrate how \thmref{thm:composition} can be 
applied, we give a simple proof of the $\Omega(\sqrt{n})$ lower bound due to
Barnum and Saks \cite{BS04} on the bounded-error quantum query complexity of
read-once functions.


\begin{corollary}[Barnum-Saks]
Let $h$ be a read-once Boolean function with $n$ variables.  Then 
$Q_2(f) = \Omega(\sqrt{n})$.  
\label{cor:bs}
\end{corollary}

\begin{proof}
We prove by induction on the number of variables $n$ that 
$\ADV(f) \ge \sqrt{n}$.  If $n=1$ then the formula is either $x_i$ or
$\neg x_i$ and taking $\Gamma=1$ shows the adversary bound is at least 1. 

Now assume the induction hypothesis holds for read-once formulas on 
$n$ variables, and let $h$ be given by a read-once formula with $n+1$ variables.
As usual, we can push any NOT gates down to the leaves, and assume that the
root gate in the formula for $h$ is labeled either by an AND gate or an OR gate.
Assume it is AND---the other case follows similarly.  In this case, $h$ can be 
written as 
$h=g_1 \wedge g_2$ where $g_1$ is a read-once function on $n_1 \le n$ bits 
and $g_2$ is a read-once function on $n_2 \le n$ bits, where $n_1+n_2=n+1$.  
We want to calculate $\adv_{\vec 1}(h)$.  Applying \thmref{thm:composition}, 
we proceed to first calculate $\beta_1=\adv(g_1)$ and $\beta_2=\adv(g_2)$.  
By the induction hypothesis, we know $\beta_1 \ge \sqrt{n_1}$ and 
$\beta_2 \ge \sqrt{n_2}$.  
We now proceed to calculate $\adv_{\vec 1}(h)=\adv_{(\beta_1,\beta_2)}(\AND)$.
We set up our AND adversary matrix as follows:
\begin{center}
\begin{tabular}{c|cccc}
   & 00 & 01 & 10 & 11 \\ \hline
00 & 0  & 0  & 0  & 0  \\ 
01 & 0  & 0  & 0  & $\beta_1$  \\
10 & 0  & 0  & 0  & $\beta_2$  \\
11 & 0  & $\beta_1$  & $\beta_2$  & 0  \\
\end{tabular}
\end{center}
This matrix has spectral norm $\sqrt{\beta_1^2 + \beta_2^2}$, and 
$\sn{\Gamma \circ D_1}=\beta_1$, and
$\sn{\Gamma \circ D_2}=\beta_2$.  Thus 
$$
\beta_1 \frac{\sn{\Gamma}}{\sn{\Gamma \circ D_1}}=
\beta_2 \frac{\sn{\Gamma}}{\sn{\Gamma \circ D_2}}=
\sqrt{\beta_1^2 + \beta_2^2}\ge \sqrt{n+1}.
$$
\end{proof}

We prove \thmref{thm:composition} in two parts.  
Our main technical lemma is given in \secref{sec:proof}, where we show a 
general result about the behavior of the spectral norm under composition of 
adversary matrices; 
we use this lemma in \secref{sec:lower} to show the lower bound
$\ADV_\alpha(h) \ge \ADV_\beta(f)$. 
This lower bound is the only direction which is needed in \corref{cor:bs}, thus
a self-contained proof of this result can be obtained by reading 
\secref{sec:proof} and \secref{sec:lower}.  
In \secref{sec:upper} we prove the upper bound 
$\ADV_\alpha(h) \le \ADV_\beta(f)$.  This is done by dualizing the spectral
norm expression for $\adv_\alpha$ and showing how the dual solutions compose.

\section{Spectral norm of a composition matrix}
\label{sec:proof}
In this section we prove our main technical lemma.  Given an adversary 
matrix $\Gamma_f$ realizing the adversary bound for $f$ and adversary
matrices $\Gamma_{g_i}$ realizing the adversary bound for $g_i$ where 
$i=1, \ldots, k$, we build an adversary matrix $\Gamma_h$
for the function $h=f \circ (g_1, \ldots, g_k)$.  The main lemma expresses
the spectral norm of this $\Gamma_h$ in terms of the spectral norms of 
$\Gamma_f$ and $\Gamma_{g_i}$. 

Let $\Gf$ be an adversary matrix for $f$, i.e.\ a matrix satisfying 
$\Gf[x,y]=0$ if
$f(x)=f(y)$, and let $\df$ be a prinicipal eigenvector of $\Gf$ with unit norm. 
Similarly, let $\Gg{i}$ be a spectral matrix for
$g_i$ and let $\dg{i}$ be a principal eigenvector of unit norm, for every
$i=1, \ldots, k$.

It is helpful to visualize an adversary matrix in the following way.  Let
$X_{f}=f^{-1}(0)$ and $Y_{f}=f^{-1}(1)$.  We order the the rows first by 
elements from $X_{f}$ and then by elements of $Y_{f}$.  In this way, the 
matrix has the following form:
$$
\Gf=
\left[
\begin{array}{cc}
0 & \Gf^{(0,1)}  \\ 
\Gf^{(1,0)} & 0   \\
\end{array}\right]
$$
where $\Gf^{(0,1)}$ is the submatrix of $\Gf$ with rows labeled 
from $X_f$ and columns labeled from $Y_f$ and $\Gf^{(1,0)}$ is the 
conjugate transpose of $\Gf^{(0,1)}$.

Thus one can see that an adversary matrix for a Boolean function corresponds 
to a (weighted) bipartite graph
where the two color classes are the domains where the function takes 
the values $0$ and $1$.  For 
$b \in \01$ let $\dgb{i}{b}[x]=\dg{i}[x]$ if 
$g_i(x)=b$ and $\dgb{i}{b}[x]=0$ otherwise.  In other words, $\dgb{i}{b}$ is 
the vector $\dg{i}$ restricted to the color class $b$.  

Before we define our composition matrix, we need one more piece of notation.
Let $\Gf^{(0,0)}=\sn{\Gf}I_{|X_f|}$, where $I$ is a $|X_f|$-by-$|X_f|$ identity
matrix and $\Gf^{(1,1)}=\sn{\Gf}I_{|Y_f|}$.

We are now ready to define the matrix $\Gh$: 
\begin{definition}
$
\Gh[x,y]=\Gf[\tx,\ty] \cdot 
\left( \bigotimes_i \Gg{i}^{(\tx_i, \ty_i)} \right) [x,y]
$
\end{definition}
A similar construction of $\Gh$ is used by Ambainis to establish the 
composition theorem for iterated functions.  

Before going into the proof, we look at a simple estimate of the spectral norm
of $\Gh$.  Notice that for any values $b_0,b_1 \in \01$ the matrix 
$\Gg{i}^{(b_0,b_1)}$ is a submatrix of 
$\Gg{i} + \sn{\Gg{i}} I$.  Thus the matrix $\Gh$ is a submatrix of the 
matrix
$$
\Gf \otimes \left( \bigotimes (\Gg{i}+\sn{\Gg{i}}I) \right).
$$
Therefore the spectral norm of $\Gh$ is upper bounded by the spectral norm of 
this tensor product matrix.
Since $\sn{\Gg{i}+\sn{\Gg{i}}\;I}=2 \sn{\Gg{i}}$ it follows that
$\sn{\Gh} \le \sn{\Gf} \cdot 2^k \prod_{i=1}^k \sn{\Gg{i}}$.  

By exploiting the block structure of $\Gh$ and the fact that 
$\Gh$ is nonnegative, we are able to prove the following tight bound, 
the key to our adversary composition theorem.

\begin{lemma}
Let $\Gh$ be defined as above for a nonnegative adversary matrix $\Gf$.
Then $\sn{\Gh}=\sn{\Gf}\cdot \prod_{i=1}^k \sn{\Gg i}$ and a principal 
eigenvector of $\Gh$ is 
$\dh[x]=\df[\tx] \cdot \prod_{i=1}^k \dgb{i}{\tx_i}[x]$.
\label{lem:spnorm}
\end{lemma}

\begin{proof}
First we will show $\sn{\Gh} \le \sn{\Gf} \cdot \prod_{i=1}^k \sn{\Gg i}$ 
by giving an upper bound on $u^* \Gh u$ for an arbitrary unit vector $u$.  

For $a \in \01^k$ let $X_a=\{x \in \01^n: \tx=a\}$.  The $2^k$ many 
(possibly empty) sets $X_a$ partition $X$.  Let $u_a$ be the vector $u$ 
restricted to $X_a$, that is $u_a[x]=u[x]$ if $x\in X_a$ and $u_a[x]=0$ 
otherwise.  The sets $\{X_a\}_{a \in \01^k}$ give rise
to a partition of the matrix $\Gh$ into $2^{2k}$ many blocks, where block
$(a,b)$ is labelled by rows from $X_a$ and columns from $X_b$.  
The $(a,b)$ block of $\Gh$ is equal to the matrix 
$\Gf[a,b] \cdot \otimes_{i=1}^k \Gg{i}^{(a_i, b_i)}$. 

Now we have
$$
u^* \Gh u = \sum_{a,b} \Gf[a,b] \cdot
	\sum_{\innersum} \left( \bigotimes_{i=1}^k \Gg{i}^{(a_i, b_i)}
       \right)[x,y] \cdot u[x]u[y].
$$
Notice that for fixed $a,b$ the inner sum is over the tensor product
$\otimes \Gg{i}^{(a_i, b_i)}$.  The largest eigenvalue of this matrix
is $\prod_{i=1}^k \sn{\Gg{i}}$, as $\sn{\Gg{i}^{(0,0)}}
=\sn{\Gg{i}^{(0,1)}}
=\sn{\Gg{i}^{(1,0)}}
=\sn{\Gg{i}^{(1,1)}}
=\sn{\Gg{i}}$.
It follows that,
$$
\sum_{\innersum} \left( \bigotimes_{i=1}^k \Gg{i}^{(a_i, b_i)}\right)[x,y]
\cdot u[x]u[y]
\le \prod_{i=1}^k \sn{\Gg i} \cdot \norm{u_{a}} \norm{u_{b}}.
$$

By the nonnegativity of $\Gf$,
\begin{eqnarray*}
u^* \Gh u &\le& \prod_{i=1}^k \sn{\Gg i} \cdot 
\sum_{a,b} \Gf[a,b] \cdot \norm{u_a} \norm{u_b} \\
          &\le& \prod_{i=1}^k \sn{\Gg i} \cdot \sn{\Gf} \cdot \norm{u}^2 \\
          &=& \sn{\Gf} \cdot \prod_{i=1}^k \sn{\Gg i}.
\end{eqnarray*}

We now turn to the lower bound.  We wish to show that  
$$
\dh[x]=\df[\tx] \cdot (\otimes_{i=1}^k \dgb{i}{\tx_i})[x]
$$
is an eigenvector of $\Gh$ with eigenvalue $\sn{\Gf}\cdot \prod_i \sn{\Gg{i}}$.

As $\Gg{i}$ is bipartite, notice that 
$\Gg{i} \dgb{i}{b}=\sn{\Gg{i}} \dgb{i}{1-b}$, for $b \in \01$.  As 
$\dg{i}$ is a unit vector it follows that 
$\norm{\dgb{i}{0}}^2=\norm{\dgb{i}{1}}^2=1/2$.  Thus 
$\norm{\otimes \dgb{i}{a_i}}^2= \prod \norm{\dgb{i}{a_i}}=1/2^k$, for any 
$a \in \01^k$.  Hence also $\norm{\dh}^2=1/2^k$.

Consider the sum
\begin{equation}
\dh^* \ \Gh \dh = \sum_{a,b} (\df[a] \cdot \otimes \dgb{i}{a_i})^* \ 
(\Gf[a,b] \cdot \bigotimes \Gg{i}^{(a_i, b_i)})
(\df[b] \cdot \otimes \dgb{i}{b_i}).
\label{sum}
\end{equation}
Notice that for fixed $a,b \in \01^k$
\begin{equation}
(\otimes \dgb{i}{a_i})^* \; \bigotimes \Gg{i}^{(a_i, b_i)} \;
(\otimes \dgb{i}{b_i}) = \prod_{i=1}^k \sn{\Gg{i}} \cdot
\norm{\otimes \dgb{i}{a_i}}^2=\frac{1}{2^k}\prod_{i=1}^k \sn{\Gg{i}}.
\label{inner}
\end{equation}
To see this, consider the two cases $a_i=b_i$ and $a_i=1-b_i$:
\begin{itemize}
  \item if $a_i=b_i$ then $\dgb{i}{a_i}=\dgb{i}{b_i}$ and 
  $\Gg{i}^{(a_i, b_i)}=\sn{\Gg{i}}I$, thus
  $(\dgb{i}{a_i})^* \Gg{i}^{(a_i,b_i)} \dgb{i}{b_i}=
    \sn{\Gg{i}} \norm{\dgb{i}{a_i}}^2$,

  \item if $a_i=1-b_i$ then $\Gg{i}^{(a_i, b_i)}$ sends $\dgb{i}{b_i}$ to
$\sn{\Gg{i}} \dgb{i}{a_i}$ and so
  $(\dgb{i}{a_i})^* \Gg{i}^{(a_i,b_i)} \dgb{i}{b_i}= 
\sn{\Gg{i}} \norm{\dgb{i}{a_i}}^2$.
\end{itemize}
\goodbreak
Substituting expression~(\ref{inner}) into the sum~(\ref{sum}) we have
\begin{eqnarray*}
\dh^* \Gh \dh &=& 
\frac{1}{2^k} \prod_{i=1}^k \sn{\Gg{i}} \sum_{a,b} \df[a]^* \Gf[a,b] \df[b] \\
&=& \frac{1}{2^k} \prod_{i=1}^k \sn{\Gg{i}} \cdot \sn{\Gf} \cdot \norm{\df}^2 \\
&=& \sn{\Gf} \cdot \prod_{i=1}^k \sn{\Gg{i}} \cdot \norm{\dh}^2.
\end{eqnarray*}
\end{proof}

\section{Composition lower bound}
\label{sec:lower}
With \lemref{lem:spnorm} in hand, it is a relatively easy matter to show 
a lower bound on the adversary value of the composed function $h$.
\begin{lemma}
  \label{lem:compose-sa}
$\ADV_\alpha(h) \ge \ADV_\beta(f)$.
\end{lemma}

\begin{proof}
Due to the maximization over all matrices $\Gamma$, the spectral bound of the
composite function $h$ is at least $\ADV_\alpha(h) \ge \min_{\ell=1}^n (
\alpha_\ell \sn{\Gh} / \sublam \Gh \ell )$, where $\Gh$ is defined as in
\lemref{lem:spnorm}.  We compute $\sublam \Gh \ell$ for $\ell = 1, \dots, n$.
Let the $\ell$-th input bit be the $q$-th bit in the $p$-th block.  Recall
that
\begin{align*}
\Gh[x,y]
  &= \Gf[\tx,\ty] \cdot \prod_{i=1}^k \Gg{i}^{(\tx_i, \ty_i)}[x^i,y^i]. \\
\noalign{We prove that} \\
(\Gh \circ D_\ell)[x,y]
  &= (\Gf \circ D_p) [\tx,\ty]
  \cdot (\Gg{p}\circ D_q)^{(\tx_p, \ty_p)}[x^p,y^p] 
  \cdot \prod_{i \ne p} \Gg{i}^{(\tx_i, \ty_i)} [x^i,y^i].
\end{align*}

If $x_\ell \ne y_\ell$ and $\tx_p \ne \ty_p$ then both sides are equal because
all multiplications by $D_p,D_q,D_\ell$ are multiplications by 1.  If this
is not the case---that is, if $x_\ell = y_\ell$ or $\tx_p = \ty_p$---then both 
sides are zero.  We see this by means of two cases:

\begin{enumerate}
  \item $x_\ell = y_\ell$: In this case the left hand side is zero due to 
  $(\Gh \circ D_\ell) [x,y] = 0$.  The right hand side is also zero because
    \begin{enumerate}
      \item if $\tx_p = \ty_p$ then the right hand side is zero as 
      $(\Gf \circ D_p) [\tx,\ty] = 0$.
      \item else if $\tx_p \ne \ty_p$ then the right hand side is zero as
      $(\Gg p \circ D_q) [x^p, y^p] = 0$.
    \end{enumerate}
  \item $x_\ell \ne y_\ell$, $\tx_p = \ty_p$: The left side is zero because
    $\Gg{p}^{(\tx_p, \ty_p)}[x^p,y^p]=
    \sn{\Gg{p}}I[x^p,y^p]=0$ since $x^p \ne y^p$.  
    The right side is also zero due to $(\Gf \circ D_p) [\tx,\ty] = 0$.
\end{enumerate}

Since $\Gh \circ D_\ell$ has the same structure as $\Gh$, by
\lemref{lem:spnorm}, $\sublam \Gh \ell = \sublam \Gf p \cdot \sublam
{\Gg p} q \cdot \prod_{i \ne p} \sn{\Gg i}$.  By dividing the two
spectral norms,
\begin{equation}
{\sn{\Gh} \over \sublam \Gh \ell}
  = {\sn{\Gf} \over \sublam \Gf p}
  \cdot {\sn{\Gg p} \over \sublam {\Gg p} q}.
\label{eq:lg/lgi}
\end{equation}

Since the spectral adversary maximizes over all adversary matrices, we 
conclude that
\begin{align*}
\ADV_\alpha(h)
&\ge \min_{\ell=1}^n {\sn{\Gh} \over \sublam \Gh \ell} \cdot \alpha_\ell \\
&= \min_{i=1}^k {\sn{\Gf} \over \sublam \Gf i}
  \cdot \min_{j=1}^{n_i} {\sn{\Gg i} \over \sublam {\Gg i} {j}}
  \cdot \alpha^i_j \\
&= \min_{i=1}^k {\sn{\Gf} \over \sublam \Gf i}
  \cdot \ADV_{\alpha^i}(g_i) \\
&= \min_{i=1}^k {\sn{\Gf} \over \sublam \Gf i } \cdot \beta_i \\
&= \ADV_\beta(f),
\end{align*}
which we had to prove.
\end{proof}

\section{Composition upper bound}
\label{sec:upper}
In this section we prove the upper bound $\ADV_\alpha(h) \le \ADV_\beta(f)$.
We apply the duality theory of semidefinite programming to obtain an 
equivalent expression for $\ADV_\alpha$ in terms of a minimization problem.
We then upper bound $\ADV_\alpha(h)$ by showing how to compose solutions 
to the minimization problems.

\begin{definition}
Let $f: S \to \01$ be a partial boolean function, where $S \subseteq
\01^n$, and let $\alpha \in \R_{+}^n$.
The \emph{minimax bound of $f$ with costs $\alpha$} is
\[
\MM_\alpha(f) = \min_p \max_{\fxnefy f} 
  \frac{1}{\sum_{\xineyi i} \sqrt{ p_x(i) p_y(i) } / \alpha_i},
\]
where $p: S \times [n] \to [0,1]$ ranges over all sets of $|S|$
probability distributions over input bits, that is, $p_x(i) \ge 0$ and
$\sum_i p_x(i) = 1$ for every $x \in S$.
\end{definition}

This definition is a natural generalization of the minimax bound introduced 
in \cite{LM04, SS06}.  As \cite{SS06} show that the minimax bound is equal to
the spectral norm formulation of the adversary method, one can similarly show
that the versions of these methods with costs are equal.

\begin{theorem}
[Duality of adversary bounds]
\label{thm:equal}
For every $f: \01^n \to \01$ and $\alpha \in \R_{+}^n$,
\begin{equation*}
\ADV_\alpha(f) = \MM_\alpha(f).
\end{equation*}
\end{theorem}

\begin{proof}[Sketch of proof.]
We start with the minimax bound with costs, substitute $q_x(i) p_x(i) / \alpha_i$, and rewrite the condition $\sum_i p_x(i) = 1$ into
$\sum_i \alpha_i q_x(i) = 1$.  Using similar arguments as
in~\cite{SS06}, we rewrite the bound as a semidefinite program,
compute its dual, and after a few simplifications, get the spectral
bound with costs.
\end{proof}

\begin{lemma}
  \label{lem:compose-mm}
$\ADV_\alpha(h) \le \ADV_\beta(f)$.
\end{lemma}

\begin{proof}
Let $p^f$ and $p^{g_i}$ for $i=1, \dots, k$ be optimal sets of probability
distributions achieving the minimax bounds.  Thus using \thmref{thm:equal}
we have
\begin{align*}
\ADV_\beta(f) &= \max_{\fxnefy f} {1 \over \sum_{\xineyi i}
	\sqrt{ p^f_x(i) p^f_y(i) } / \beta_i }, \\
\ADV_{\alpha^i}(g_i) &= \max_{\fxnefy {g_i}} {1 \over \sum_{\xineyi j}
	\sqrt{ p^{g_i}_x(j) p^{g_i}_y(j) } / \alpha^i_j}. \\
\end{align*}
Define the set of probability distributions $p^h$ as 
$p^h_x(\ell) = p^f_\tx(i) p^{g_i}_{x^i}(j)$, where the $\ell$-th input bit is the
$j$-th bit in the $i$-th block.  This construction was first used by Laplante,
Lee, and Szegedy~\cite{LLS06}.  We claim that $p^h$ witnesses that
$\ADV_\alpha(h) \le \ADV_\beta(f)$: 
\begin{align*}
\ADV_\alpha(h)
&\le \max_{\fxnefy h} {1 \over \sum_{\xineyi \ell} \sqrt{ p^h_x(\ell) 
p^h_y(\ell) } / \alpha_\ell } \\
&= 1 \Bigg/ \min_{\fxnefy h} \sum_{\xineyi \ell} \sqrt{ p^f_\tx(i) p^f_\ty(i) }
	\sqrt{ p^{g_i}_{x^i}(j) p^{g_i}_{y^i}(j) }
	/ \alpha^i_j \\
&= 1 \Bigg/ \min_{\tx, \ty \atop f(\tx) \ne f(\ty)}
	\sum_i \sqrt{ p^f_\tx(i) p^f_\ty(i) }
	\min_{x^i, y^i \atop {g_i(x^i)=\tx_i \atop g_i(y^i)=\ty_i}}
	\sum_{j: x^i_j \ne y^i_j} \sqrt{ p^{g_i}_{x^i}(j) p^{g_i}_{y^i}(j) } 
	/ \alpha^i_j \\
&\le 1 \Bigg/ \min_{\tx, \ty \atop f(\tx) \ne f(\ty)}
	\sum_{i: \tx_i \ne \ty_i} \sqrt{ p^f_\tx(i) p^f_\ty(i) }
	\min_{x^i, y^i \atop g_i(x^i) \ne g_i(y^i)}
	\sum_{j: x^i_j \ne y^i_j} \sqrt{ p^{g_i}_{x^i}(j) p^{g_i}_{y^i}(j) } 
	/ \alpha^i_j \\
&= 1 \Bigg/ \min_{\tx, \ty \atop f(\tx) \ne f(\ty)}
	\sum_{i: \tx_i \ne \ty_i} \sqrt{ p^f_\tx(i) p^f_\ty(i) }
	\ /\  \ADV_{\alpha^i}(g_i) \\
&= 1 \Bigg/ \min_{\tx, \ty \atop f(\tx) \ne f(\ty)}
	\sum_{i: \tx_i \ne \ty_i} \sqrt{ p^f_\tx(i) p^f_\ty(i) }
	\ /\  \beta_i \\
&= \ADV_\beta(f),
\end{align*}
where the second inequality follows from that fact that we have
removed $i: \tx_i = \ty_i$ from the sum and the last equality follows from
\thmref{thm:equal}.  
\end{proof}

Laplante, Lee, and Szegedy~\cite{LLS06} proved a similar bound in a stronger
setting where the sub-functions $g_i$ can act on the same input bits.  They
did not allow costs of input bits.  This setting is, however, not applicable
to us, because we cannot prove a matching lower bound for $\ADV_\alpha(h)$.

\section*{Acknowledgements}

We thank Mehdi Mhalla for many fruitful discussions.


\newcommand{\slasha}{\discretionary{/}{/}{/}}

\end{document}